\documentclass[angeo]{copernicus}

\usepackage{color}

\begin{document}

\title{Thin current sheets caused by plasma flow gradients in space and astrophysical plasma}
\author[1,2]{Dieter Nickeler}
\author[2]{Thomas Wiegelmann}
\affil[1]{Astronomical Institute AV \v{C}R Ond\v{r}ejov, Fri\v{c}ova 298, 25165 Ondrejov, Czech Republic}
\affil[2]{Max-Planck-Institut f\"ur Sonnensystemforschung,
Max-Planck-Strasse 2, 37191 Katlenburg-Lindau, Germany}


\runningtitle{MHD with plasma flow}

\runningauthor{Nickeler et al.}

\correspondence{Nickeler\\ (EMAIL: nickeler@asu.cas.cz)}

\received{}
\pubdiscuss{} 
\revised{}
\accepted{}
\published{}


\firstpage{1}

\maketitle
\begin{abstract}
Strong gradients in plasma flows play a major role in space and astrophysical plasmas.
A typical situation is that a static plasma equilibrium is surrounded by a
plasma flow, which can lead to strong plasma flow gradients at the separatrices
between field lines with different magnetic topologies, e.g.,
planetary magnetospheres, helmet streamers in the solar corona,
or at the boundary between the heliosphere
and interstellar medium. Within this work we make a first step to understand
the influence of these flows towards the occurrence of current sheets in a
stationary state situation. We concentrate here on incompressible
plasma flows and 2D equilibria, which allow us to find analytic solutions
of the stationary magnetohydrodynamics equations (SMHD).
First we solve
the magnetohydrostatic (MHS) equations with the help of a
Grad-Shafranov equation and then we transform these static equilibria into
a stationary state with plasma flow. We are in particular interested
to study SMHD-equilibria with strong plasma flow gradients
perpendicular to separatrices.
We find that induced thin current sheets occur naturally in
such situations. The strength of the induced currents
depend on the Alfv\'en Mach number and its gradient, and on the magnetic
field.
\end{abstract}
%
\introduction
Plasma flows around separatrices play an important role
in many astrophysical and space plasmas.
Significant flows occur thereby mainly on open field lines,
while the plasma on closed field lines is approximately at rest.
One example is the magnetosphere surrounded by the solar wind flow,
where both regions are separated by the magnetopause. Another example
are coronal helmet streamers, where the closed arcade type magnetic
structures are surrounded by open magnetic field lines on which the
stationary solar wind is streaming.

A third example, but with different constraints,
is the region far away from a star, which is embedded in the
counterflowing interstellar medium. Due to the interaction between the stellar wind, e.g., the
solar wind, and the counterflowing interstellar medium, a separatrix forms, separating
the \lq inner\rq~stellar wind from the
\lq outer\rq~interstellar medium: the domain inside is called
an astrosphere (heliosphere for the sun), the corresponding separatrix is called
an astropause (heliopause). This is a similar
situation as described for the helmet streamers in the paragraph before, namely
different flow regimes inside and outside of some boundary layer
(\lq inner\rq~and \lq outer\rq~field lines). In the case of
astrospheres/of the heliosphere, however,
the scenario implies a structure with almost completely open field lines.
Additionally, the flow is
non-zero also on the \lq inner\rq~field lines, but shows a strong gradient
\citep[see, e.g.,][for details]{baranov:etal70,baranov:etal71,nickeler:etal06a}.

These situations where regions with and without plasma flow are
separated by rather thin boundary layers necessarily lead to
strong flow gradients in these layers. Within this work we
aim to study the relation of these flow gradients to current sheets.
Thin current sheets are important, because due to current driven micro-instabilities
a fine resistivity occurs in these regions and the usual assumption of an ideal
conducting space plasma breaks down. Consequently resistive plasma instabilities
like magnetic reconnection can occur as a consequence of current sheets and
when additional sufficient free energy for eruptions is available in the configuration
\citep[see for example][]{schindler:etal93}. Such processes occur, e.g., as substorms
in magnetospheres and flares in the solar corona, \citep[see][for a recent study of
similarities and difference between substorms and flares]{birn:etal09}.
\citet{wiegelmann:etal95}, and \citet{becker:etal01} studied the formation of thin
current sheets as a
sequence of quasi-static magnetotail equilibria. Quasi-static means that dynamical effects
and the influence of plasma flow can be neglected compared to magnetic forces and pressure
gradients. This assumption is well fulfilled in the magnetosphere during quiet times
\citep[][]{schindler:etal82} and the strongest currents form in the center of the magnetospheric
plasma sheet. The influence of a field-aligned parallel plasma flow in magnetospheres has
been studied for example in \citet{birn91}
and for solar MHD-equilibria
in \citet{wiegelmann:etal98}, \citet{petrie:etal99}, \citet{petrie:etal02},
and \citet{petrie:etal05}.
A significant influence of a {\it smooth}
plasma flow itself onto the magnetic field and plasma configuration is
rather low for flow speed well below the Alfv\'en speed $v \ll v_A$. Far less studied
has been the influence of significant small scale gradients in the plasma flow, which
is the topic of this paper. We are in particular interested to investigate to
which extend thin current sheets at boundary layers can be associated with corresponding
gradients in the plasma flow.
We outline the paper as follows. In section \ref{basics} we present the basic equations,
outline how static equilibria can be transformed into stationary ones, and discuss
the relation between sharp flow gradients and the occurrence of thin current sheets.
Sections \ref{app1} and \ref{app2} contain applications to solar coronal and
magnetospheric structures, respectively. Finally we summarize our results
in section \ref{conclusions}.
\section{Basic assumptions and equations}
\label{basics}
We apply the theory of ideal MHD. This is justified, because of the high conductivity
in many space and astrophysical plasmas.
For simplicity and to concentrate on essential \lq flow\rq~effects,
we restrict our research to configurations
with an incompressible plasma flow. We are interested in the physical effects that
occur for large gradients of the Alfv\'en Mach number perpendicular to the field lines.
We are aware that compressible effects
might become important for some space plasma applications, in particular for
stratified plasmas like the solar corona and chromosphere, especially
on large scales, see, e.g. \citet{petrie:etal02} or \citet{petrie:etal05}.

For stationary, ideal and incompressible MHD we have to solve the following equations:
mass continuity equation (\ref{mce}),
the Euler or momentum equation with isotropic pressure $P$ (\ref{ee}), the
stationary induction equation including the ideal Ohm's law (\ref{ie}), Amp\`ere's law
(\ref{al}), the solenoidal condition for the magnetic
field (\ref{sc}), and the condition for incompressibility (\ref{ice}):
\begin{eqnarray}
  \vec\nabla\cdot\left( \rho\vec{\rm v} \right) &=& 0\, ,\label{mce}\\
   \rho\left( \vec{\rm v}\cdot\vec\nabla\right)\vec{\rm v} &=& \vec
j\times\vec B -
\vec\nabla P\, \label{ee},\\
    \vec\nabla\times\left(\vec{\rm v}\times\vec B\right) &=&\vec 0\,
,\label{ie}\\
     \vec\nabla\times\vec B &=& \mu_{0}\vec j\, ,\label{al}\\
      \vec\nabla\cdot\vec B &=& 0\, ,\label{sc}\\
       \vec\nabla\cdot\vec{\rm v} &=& 0\,\label{ice} ,
        \end{eqnarray}
where $\rho$ is the mass density, $\vec{\rm v}$ the plasma velocity, $\vec B$ the magnetic field,
$\vec j$ the current density vector, and $P$ the thermal or plasma pressure.

Due to the incompressibility, the mass continuity equation can be written as
$\vec {\rm v}\cdot\vec\nabla\rho=0$, so that the density is constant on
streamlines.
We now introduce the auxilliary flow vector or streaming vector $\vec
w:=\sqrt{\rho}\,\vec{\rm v}$. With the Bernoulli pressure, defined by
$\Pi:=P+\frac{1}{2}\vec w ^2$, we can rewrite the above equations as
  \begin{eqnarray}
   \vec\nabla\cdot\vec w &=& 0\, ,\label{divwfree}\\
   \frac{1}{\mu_{0}}\left( \vec\nabla\times\vec
B\right)\times\vec B - \left( \vec\nabla\times\vec w \right)\times\vec w
&=& \vec\nabla \Pi \, ,\label{equ}\\
     \vec\nabla\times\left(\frac{1}{\sqrt{\rho}}\,\vec w\times\vec B\right) &=&
\vec 0\, ,\label{indeq} \\
      \vec\nabla\cdot\vec B &=& 0\, . \label{solenoid}
       \end{eqnarray}
The momentum equation Eq.\,(\ref{equ}) is written such that the analogy
with magnetohydrostatic equilibria (MHS), given by
\begin{equation}
\mu_{0}^{-1}\left( \vec\nabla\times\vec B\right)\times\vec B=\vec\nabla P\, ,
\end{equation}
is obvious.

The assumption of a field-aligned flow enhances the probability that
the flow is stable in the frame of ideal MHD, \citep[see the discussion in][]{hameiri98}.
The assumption of $\vec{\rm v}\times\vec B=0$ leads to a vanishing
electric field in ideal MHD. This can be seen with the help
of the uncurled induction equation, Eq.\,(\ref{ie}), which is basically ideal
Ohm's law $\vec E+\vec{\rm v}\times\vec B=\vec 0$ ($\vec E=$electric field).
Therefore the stationary ideal Ohm's law  (\ref{indeq}) is fulfilled identically.
Under these assumptions the set of equations reduces to:
   \begin{eqnarray}
   && \vec B\cdot\vec\nabla M_{A} = 0\, ,\label{solenoid3}\\
   &&  \vec\nabla \Pi  =\frac{\left(1-M_{A}^2\right)\left(
     \vec\nabla\times\vec B\right)\times\vec B}{\mu_{0}} - \frac{|\vec B|^2}{2\mu_{0}}
     \,\vec\nabla\left( 1-M_{A}^2\right)\, ,
     \label{equ2}\\
     && \vec\nabla\cdot\vec B = 0\, , \label{solenoid4}
       \end{eqnarray}
where $M_{A}$ is the Alfv\'en Mach number, defined via
\begin{equation}
\vec w\equiv\pm M_{A}\vec B/\sqrt{\mu_{0}}\,\, ,
\end{equation}
and where the $\pm$ indicates that if the pair $(\vec w, \vec B)$ is a solution of
the Eqs.\,(\ref{equ}) and (\ref{equ2}), then $(-\vec w, \vec B)$ is also a solution, and
basically also $(\vec w, -\vec B)$ and $(-\vec w, -\vec B)$. We will mainly parameterize
this behaviour by $M_{A}$ and by $\vec B$, starting from the viewpoint
of MHS theory, as the form (\ref{equ2}) of the incompressible SMHD
equations allows to derive transformation equations which transform MHS
equilibria into stationary ones
\citep[see][for the mathematical details.]{gebhardt:etal92,nickeler:etal06a}.
\subsection{2D stationary states}
The theory explained so far is general. In the following we concentrate
on configurations with one axis of symmetry, e.g. the $z$-axis in a cartesian
coordinate system $x,y,z$. Consequently all quantities are functions of
$x$ and $y$ only and we can solve the solenoidal equation (\ref{solenoid4}) by
introducing a flux function $\alpha(x,y)$ of the form
$\vec B=\vec\nabla\alpha\times\vec e_{z}$.
This reduces the stationary incompressible equations in 2D to
\begin{eqnarray}
  M_{A} &=& M_{A}(\alpha)\, ,\label{solenoid3new}\\
  \vec\nabla \Pi  &=&-\frac{\left(1-M_{A}^2\right)
   \Delta\alpha\vec\nabla\alpha\,}{\mu_{0}} - \frac{|\vec\nabla\alpha|^2}{2\mu_{0}}
    \,\vec\nabla\left( 1-M_{A}^2\right)\, .
     \label{equ2new}
\end{eqnarray}
Now we perform a transformation by assuming that $\alpha(x,y)$ is a function of
another \lq flux function\rq $A(x,y)$, i.e. $\alpha=\alpha(A)$, such that
the {\it stationary} equation, Eq.\,(\ref{equ2new}), reduces to a form of the
equation mathematically similar to the Grad-Shafranov equation
\begin{equation}
\quad\frac{d P_{MHS}}{dA}= -\frac{1}{\mu_{0}}\Delta A \, ,
\label{gsg0}
\end{equation}
describing MHS equilibria.

Then the equivalence between
the MHS equation and the momentum equation of ideal, stationary but
non-static MHD with incompressible, field-aligned flow is shown.
As $\alpha$ is a function of $A$,  $M_{A}$ is also a function of
the \lq new\rq~flux function $A$. Derivatives with respect to $A$ will now and in the following
be expressed by a prime, e.g., $d\alpha/dA=\alpha'$.
With the help of the relation $M_{A}=M_{A}(A)$ we can
rewrite the Euler equation Eq.\,(\ref{equ2new})
\begin{eqnarray}
\vec\nabla \Pi  &=& -\frac{\left(1-M_{A}^2\right)\left(
\alpha''\,\left(\vec\nabla A\right)^2 + \alpha'\,\Delta A \right)
\alpha'\vec\nabla A}{\mu_{0}}\nonumber \\
&& -\frac{\alpha'^2|\vec\nabla A|^2}{2\mu_{0}}\vec\nabla\left(1-M_{A}^2\right)\nonumber
\\
&=& -\frac{\left(1-M_{A}^2\right)}{\mu_{0}} \alpha'^2\Delta A\vec\nabla A
 \nonumber \\
 && -\frac{\left(\vec\nabla A\right)^2
\vec\nabla A}{2\mu_{0}}\left[\left(1-M_{A}^2\right)
\alpha'^2\right]' \, .
\label{trafo1}
\end{eqnarray}
Let us remark that the Alfv\'en Mach number $M_A$ can be expressed as a function
of $\alpha$ or $A$, but is not restricted further.
We have therefore the freedom to choose this function arbitrary
without loss of generality. A reasonable choice to eliminate the term
$(\vec \nabla A)^2$ in (\ref{trafo1}) is
\begin{equation}
(1-M_{A}^2)\alpha'^2\equiv 1\, ,
\label{machalpha}
\end{equation}
and therefore  the Euler equation (\ref{trafo1}) simplifies to
a single partial differential equation for the new flux function $A(x,y)$
\begin{eqnarray}
\vec\nabla \Pi  = -\frac{1}{\mu_{0}}\Delta A\vec\nabla A\quad\Rightarrow
\quad\frac{d\Pi}{dA}= -\frac{1}{\mu_{0}}\Delta A . \label{gsg1}
\end{eqnarray}
In any case,  Eq.\,(\ref{gsg1}) is mathematically identical with
Eq.\,(\ref{gsg0}), but contains plasma flow. Physically
this equation reduces to the static Grad-Shafranov equation only
for the limit $M_{A}\rightarrow 0$, implying $\Pi(A)\rightarrow P(A)$.
Consequently, any solution $A(x,y)$ of the \lq MHS\rq~equation
(\ref{gsg1}) (or equivalently the MHS equation (\ref{gsg0}))
can be used to derive a solution of the stationary, incompressible MHD
by integrating  equation (\ref{machalpha}) 
\begin{eqnarray}
\alpha=\pm\int\,\frac{dA}{\sqrt{1-M_{A}(A)^2}}\, . \;\; \label{alphafromA}
\end{eqnarray}

With this form we can specify a plasma flow via the Alfv\'en Mach number
$M_A(A)$. Because $A$ is constant on magnetic field lines this is also
true for $M_A$ and $\alpha$. Physically this means that we can
specify on which field lines plasma is flowing with a certain Mach number.
It is in particular possible to calculate separatrix field lines in the
static case and specify plasma flow only on one side of this separatrix,
e.g. to model plasma flow around a static magnetosphere or helmet streamer
configuration.

Equation (\ref{alphafromA}) is also equivalent to
\begin{eqnarray}
A=\pm\int\,{\sqrt{1-M_{A}(\alpha)^2}}\,d\alpha\, . \label{Afromalpha}
\end{eqnarray}
Some care has to be taken for multi-valued functions $M_{A}(A)$ or $M_{A}(\alpha)$,
where one has to distinguish between the different branches of solutions.
This is, however, not a major problem and similar to the problem of
multi-valued functions $\Pi(A)$ in the static Grad-Shafranov theory,
which has been addressed in \cite{wiegelmann:etal98} to model triple
coronal helmet streamer configurations.
\subsubsection{Influence on the electric current density}
Inserting the ansatz $\vec B=\vec\nabla\alpha\times\vec e_{z}$ into
(\ref{al}) and by computing the Laplacian of $\alpha$ with
equation (\ref{alphafromA}) we find
the connection between the Alfv\'en Mach number and current density
\begin{eqnarray}
-\mu_0 j_{z}&=& \Delta\alpha \nonumber \\
&=&
\pm \frac{M_{A} M_{A}'}{\left(1-M_{A}^2\right)^\frac{3}{2}}\left(\vec\nabla A\right)^2
\pm \frac{1}{\sqrt{1-M_{A}^2}}\Delta A \, . \label{current1}
\end{eqnarray}
The $\pm$ sign again reflects the freedom of the transformation
relations Eqs.\,(\ref{alphafromA}) and \,(\ref{Afromalpha}) with respect to the
direction of the magnetic field and the symmetry of the Lorentz force.
 The first term in equation (\ref{current1}) corresponds to a
current induced by the plasma flow and the second part modifies
(enhances) the static equilibrium current $-\mu_{0}\, j_{z,{\rm static}}=-\Delta
A$.

For static potential fields this part of the current vanishes
also in the stationary state with flow, i.e. as $\Delta A=0$
only the first term on the right side of Eq.\,(\ref{current1}) contributes to the electric current
density.
We expand equation (\ref{current1}) for small Alfv\'en Mach numbers $M_A \ll 1$
which leads to
\begin{equation}
-\mu_0 j_{z}= \pm (M_A+\frac{3}{2} M_A^3) M_{A}' \, \left(\vec\nabla A\right)^2
\pm (1+\frac{1}{2} M_A^2) \Delta A\, ,
\end{equation}
and if we neglect all quadratic and higher terms in $M_A$, we find:
\begin{equation}
-\mu_0 j_{z}=\pm M_A M_{A}' \, \left(\vec\nabla A\right)^2
\pm  \Delta A \, .
\label{current2}
\end{equation}
Inserting the definitions of the magnetic field and equilibrium
current and using the definition $M_{A}' \, \nabla A=\frac{d M_{A}}{d
A} \, \nabla A = \nabla M_A$ (this can be done also already in
(\ref{current1}) and does not depend on the assumption of small Mach
numbers), we get from (\ref{current2})
\begin{equation}
-\mu_0 j_{z}=\pm (M_{A} \vec\nabla M_{A}\cdot\vec\nabla A  +  j_{z,{\rm static}})\, .
\label{current2a}
\end{equation}
Consequently for small Alfv\'en Mach numbers the equilibrium currents
are basically unmodified by the plasma flow and the induced currents
depend linearly on the magnetic field strength, the Alfv\'en Mach
number and the gradient of the Alfv\'en Mach number. For further
approximations on the relative strength of the equilibrium currents
and the flow induced currents we assume that the magnetic field
equilibrium does change on a length scale $l_{\rm static}$ and the
plasma flow on a scale $l_{\rm flow}$, which allows us to roughly
approximate the gradient and Laplacian:
\begin{equation}
-\mu_0 j_{z}  \approx \pm \left( M_A \frac{M_A}{l_{\rm flow}} \, B  +  \frac{B}{l_{\rm
static}} \right)
\end{equation}
So to compare the relative strength of the two contributions we have
to compare $\frac{M_A^2}{l_{\rm flow}}$ and $\frac{1}{l_{\rm
static}}$ and get as the ratio of induced and equilibrium current
$M_A^2 \frac{l_{\rm static}}{l_{\rm flow}}$. Consequently we get
(for slow plasma flows with $M_A \ll 1$) only a significantly large
induced current if the plasma flow changes on a much smaller length
scale as the typical scale of the configuration $l_{{\rm flow}} \ll
l_{{\rm static}}$. Such a situation is typically fulfilled at
boundary layers, e.g., the magnetopause or the separatrix between
open and closed field lines in coronal helmet streamers.
\section{Application to coronal helmet streamers and plasmoids}
\label{app1}
\begin{figure}
\vspace*{2mm}
\begin{center}
\includegraphics[width=7.3cm]{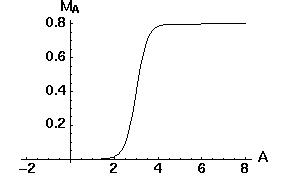} \\ 
\vspace*{5mm}
\includegraphics[width=7.3cm]{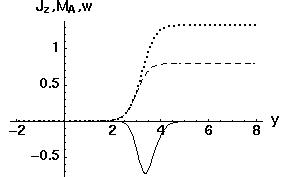} \\ 
\end{center}
\caption{Transformation from a one-dimensional homogenous current-free potential
field $A=B_0 \, y$.
Top panel:$M_A(A)$
as defined in (\ref{example_MA}),
with $M_1=0.4$, $M_2=0.8$, $d=2$, $A_c=3$
Bottom panel: $M_A(y)$ in dashed line, $w(y)$ in thick dotted line and the resulting
electric current density $j_z(y)$ in solid line.}
\label{f01}
\end{figure}

%
In the following we provide some example solutions for MHD-equilibria
with plasma flow. We construct these configurations by first solving the
MHS problem (\ref{gsg1}) and then by transforming the
resulting static flux function $A(x,y)$ into the solution of the stationary
problem $\alpha(x,y)$ with the help of equation (\ref{alphafromA}).
We prescribe the Alfv\'en Mach number as a function of $A$ in the useful form
\begin{equation}
M_A(A)=M_1+(M_2-M_1)\, \tanh \left(d (A-A_c) \right),
\label{example_MA}
\end{equation}
where $M_1$, $M_2$, $d$, and $A_{c}$
are free parameters \footnote{Please note that the Alfv\'en Mach number
can become negative for plasma flows antiparallel to the field lines.}, the scale
on which the flow changes ($d \propto 1/l_{{\rm flow}}$ is an inverse length),
and the value of the separatrix field line $A_c$. The functional form of
$M_A$ in (\ref{example_MA}) has been chosen in order to provide the strongest
flow gradient at the separatrix field line $A_c$. Figure \ref{f01} top panel
shows $M_{A}$ as a function of $A$
for $M_1=0.4$, $M_2=0.8$, $d=2$, and $A_c=3$.

As an example we apply the transformation (\ref{alphafromA}) with $M_A$ in the
form of (\ref{example_MA}) to a homogeneous potential
magnetic field $\vec B=B_0  \vec e_x$
with $B_0=1$. This is a simple 1-D equilibrium
with the static flux function $A=B_0 y$
and all quantities (both in the static and stationary case) are only a function
of $y$. Figure \ref{f01} bottom panel shows the Alfv\'en Mach number $M_A(y)$ (dashed line),
the streaming vector $w(y)$ (dotted line) and the
corresponding formation of a current sheet $j_z(y)$ at the separatrix field line $A_c=3$ (solid line).

In the following we study more sophisticated static equilibria and their transformation
to stationary incompressible MHD-equilibria.
For a better visualization we will present examples of MHS equilibria that do not
show the extremely small scale flows compared to the equilibrium current scales. We
therefore have to use larger Alfv\'en Mach numbers.
\subsection{Linear MHS equilibria}
\begin{figure}
\vspace*{2mm}
\begin{center}
\includegraphics[width=5.0cm]{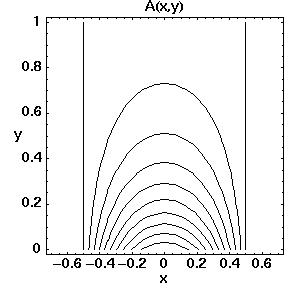} \\ 
\vspace*{3mm}
\includegraphics[width=6.0cm]{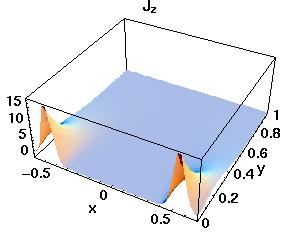} \\ 
\vspace*{3mm}
\includegraphics[width=6.0cm]{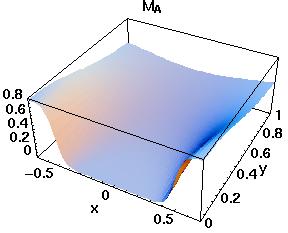} 
\end{center}
\caption{2D potential field configurations in the form
$A(x,y)=B_0 \exp(-k \pi y/L) \cos(-\sqrt{k^2-c^2} \pi x/L)$, with
$B_0=k=L=1$, $c=0.0$ and the Alfv\'en Mach number profile function
(\ref{example_MA}) with $d=5$, $M_1=0.4$, $M_2=0.0$, $A_c=0$.
This profile has a steep gradient
at the boundary between open and closed field lines.
Top panel: magnetic field lines (contour lines of $A(x,y)$),
center: formation of flow driven
current sheets $J_z(x,y)$, Bottom: Alfv\'en Mach number $M_A(x,y)$.}
\label{f02}
\end{figure}

\begin{figure}
\vspace*{2mm}
\begin{center}
\includegraphics[width=5.0cm]{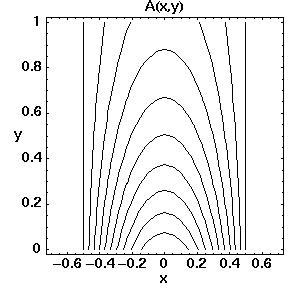} \\
\vspace*{3mm}
\includegraphics[width=6.0cm]{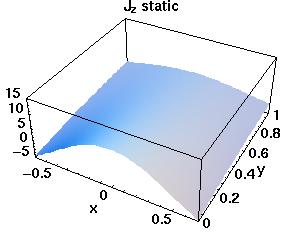} \\ 
\vspace*{3mm}
\includegraphics[width=6.0cm]{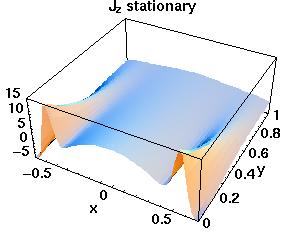} \\ 
\vspace*{3mm}
\includegraphics[width=6.0cm]{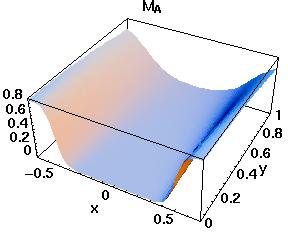} \\ 
\end{center}
\caption{Same transformation as in figure \ref{f02}
, including a linear MHS
current $J_z(A)=c \cdot A$, $c=0.9$. Second and third panel show the split
of linear (MHS, without flow) and the full current (with flow).}
\label{f03}
\end{figure}

\begin{figure}
\vspace*{2mm}
\begin{center}
\includegraphics[width=5.0cm]{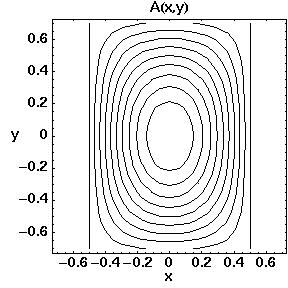} \\ 
\vspace*{3mm}
\includegraphics[width=6.0cm]{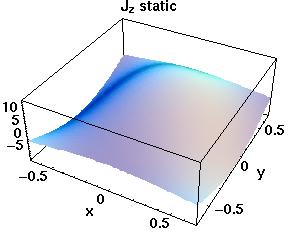} \\ 
\vspace*{3mm}
\includegraphics[width=6.0cm]{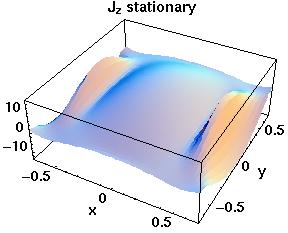} \\ 
\vspace*{3mm}
\includegraphics[width=6.0cm]{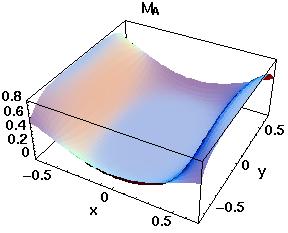} \\ 
\end{center}
\caption{Same transformation as in figure \ref{f02}, but including a 
linear MHS current in the form $J_z(A)=c \cdot A$ with $c=1.2$. Here $c > k$.}
\label{f04}
\end{figure}

To derive 2D static equilibria we solve the Grad-Shafranov equation
(\ref{gsg1}) for a linear current. Such configurations have been studied
for triple coronal helmet streamer configurations in \cite{wiegelmann98},
but here we limit our research to single helmet streamers and concentrate
on the effect of plasma flow on open field lines. A linear current means
that the function $\Pi(A)$ in equation (\ref{gsg1}) has the
form $\Pi(A)=\frac{c^2}{2} A^2$.
In this case the Grad-Shafranov equation reduces to a linear
Helmholtz equation
\begin{equation}
 -\Delta A=c^2 \, A
 \label{gsg_lin}
\end{equation}
and can be solved by separation of variables.
Let us remark that in general the static Grad-Shafranov equation
has the form $\Pi(A)=p(A)+\frac{B_z^2}{2}$, with the plasma pressure
$p$ and a magnetic shear field $B_z$ in the invariant direction.
In the case of $p(A)=0$ one obtains linear force-free configurations
and else static equilibria. In both cases the electric current
$j_z(A)=\frac{\partial \Pi}{\partial A}$ is linear in $A$. The particular
choice $c=0$ corresponds to current-free potential fields.
As solution of (\ref{gsg_lin}) one gets by separation of variables
\begin{eqnarray}
A(x,y) &=& B_0 \exp \left(-\frac{\nu \pi y}{L} \right) \cos \left(\frac{k \pi x}{L} \right)
\;\; {\rm for} \; c<k \, , \\
A(x,y) &=& B_0 \cos \left(\frac{\omega \pi y}{L} \right) \cos \left(\frac{k \pi x}{L} \right)
\;\; {\rm for} \; c>k \, , \\ \nonumber
\end{eqnarray}
with $\nu=\sqrt{k^2-c^2}$ and $\omega=\sqrt{c^2-k^2}$. Linear combinations
of these particular solutions are also solutions of the linear Helmholtz-equation
(\ref{gsg_lin}). These solutions were also studied by \cite{hood:etal90}
modeling prominence arcades and \cite{petrie06}
modeling coronal loops.
Here we consider only three particular
cases with $B_0=k=L=1$ and different values of $c$.
The top panels of figures \ref{f02}, \ref{f03}, and \ref{f04}
show magnetic field lines (equi-contour plots of the flux function
$A(x,y)$ for $c=0$, $c=0.9$, and $c=1.2$, respectively). The case $c=0$
in figure \ref{f02} corresponds to a current-free potential field.
Introducing a moderate linear current with $c<k$
leads to a stretching of the configuration (top panel in figure \ref{f03})
and a smooth electric current density distribution (second panel in figure \ref{f03}).
For the case $c>k$, as shown in the top panel of figure \ref{f04}, the magnetic
topology changes and wet get plasmoid-like configurations, which, however,
also have a smooth current density distribution in equilibrium (second panel).
%
%
\subsubsection{Transformation to stationary states}
We transform these static equilibria into stationary ones using
equation (\ref{alphafromA}) and $M_A$ in the form (\ref{example_MA})
with $d=5$, $M_1=0.0$, $M_2=0.8$, $A_c=0$.
We choose $M_{2} > M_{1}$, in order to prescribe
a plasma flow on the outside ($x=\pm 0.5$) of the configuration, where
the flux function becomes negative. Inside the streamer and plasmoid
($-0.5 < x <0.5$) the flux function is positive. The plasma flow is
chosen in order to be maximal at the open separatrix field line
and the plasma is basically at rest inside the configuration, where
the magnetic field lines are closed (bottom panels in figures
\ref{f02}, \ref{f03}, \ref{f04}).
With the current transformation equation (\ref{current1})
we compute the total current density:
In the case of the static potential field currents only
occur from the first term in (\ref{current1}) and are
driven by the gradient of the plasma flow (center panel
in figure \ref{f02}). As a consequence of the sharp
gradient in the flow, a thin current sheet forms at
the separatrix ($x=\pm 0.5$). Such a thin current sheet
forms also additional to the smooth equilibrium currents
for the linear current cases (third panels in figures
\ref{f03} and \ref{f04}). In regions with
weak or no plasma flow the equilibrium current does basically not
change, whereas a current sheet forms at the separatrix field line.
Due to the strong gradient in the plasma flow this induced
current sheet is much thinner and the current density is higher
than in the equilibrium current. The equilibrium currents are
strongest in the center of the configuration.
\section{Application to magnetospheres}
\label{app2}
%

%
%
%
\subsection{Non-linear Grad-Shafranov equation, Liouville equation}
In the static case without magnetic shear field the Grad-Shafranov-equation
(\ref{gsg1}) reduces to
\begin{equation}
\Delta A = -\mu_{0}\,\frac{\partial}{\partial A}\left(p(A) \right),  \label{gsg2}
\end{equation}
where $p(A)$ is the plasma pressure.
Under the assumption of a local thermodynamical equilibrium  the plasma pressure
function can be derived from kinetic theory in the form
\citep[see][for details]{schindler:book06}
\begin{equation}
p(A)= \frac{1}{2}\,\hat{p}\,\exp(-2A/\hat{A}) \, ,
\end{equation}
where $\hat{A}$ and $\hat{p}$ are normalization constants.
This leads to an equation in the form
\begin{equation}
\Delta A = \lambda \, \exp(-c A) \, , \label{lio1}
\end{equation}
with constants
\begin{eqnarray}
\lambda &=& \mu_{0}\,\frac{\hat{p}}{\hat{A}} \, , \\
c &=& \frac{2}{\hat{A}} \, ,
\end{eqnarray}
and the typical lengthscale $\hat{l}$, and the typical
magnetic field $\hat{B}$, defined by
\begin{eqnarray}
\lambda c &=& \left(\frac{2\hat{p}}{\hat{A}^2/\left(\hat{l}^2\mu_{0}
\right)}\right)\equiv\frac{2}{\hat{l}^2} \, , \\
\hat{B} &=& \frac{\hat{A}}{\hat{l}} \, .
\end{eqnarray}
Before we continue to compute analytical and exact 2D solutions,
we first present a well-known case in 1D, namely the Harris-sheet, to
explain how the transformation from MHS to incompressible SMHD works.
%
%
\subsection{1D Harris-sheet}
A well known, 1D equilibrium current sheet solution of the
Liouville's equation (\ref{lio1})
is the Harris-sheet \citep{harris62}.
A Harris-sheet like force-free equilibrium has recently been found
by \cite{harrison:etal09}. The Harris sheet is a 1D-solution,
where all quantities depend only on the $y$
coordinate, and is given by
\begin{equation}
A(y)=\hat{A}\,\ln\cosh(y/\hat{l})\, .
\end{equation}
The static equilibrium quantities for the Harris-sheet are
shown with solid lines in figure \ref{f05}. The top panel
shows the flux-function $A(y)$, the second panel the corresponding
magnetic field $B_x$ and the bottom panel the equilibrium
electric current density.
\subsubsection{Transformation to stationary states}
\begin{figure}
\vspace*{2mm}
\begin{center}
\includegraphics[width=7.2cm]{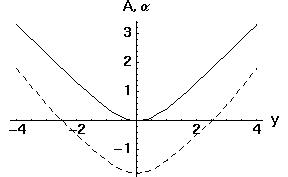} \\ 
\vspace*{5mm}
\includegraphics[width=7.2cm]{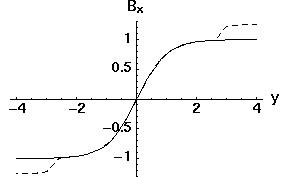} \\ 
\vspace*{5mm} 
\includegraphics[width=7.2cm]{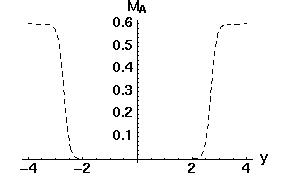} \\ 
\vspace*{5mm}
\includegraphics[width=7.2cm]{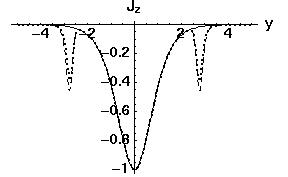} \\ 
\end{center}
\caption{
From top to bottom we show the flux-functions $A(y),\alpha(y)$,
the magnetic field $B_x(y)$, the Alfv\'en Mach number $M_A(y)$ and
the electric current density $j_z(y)$. Solid lines correspond to
the  magnetostatic case and dashed lines to stationary MHD with
the profile (\ref{example_MA}) and
$M_1=0$, $M_2=0.5$, $A_c=2$ and $d=5$.
In the bottom panel we show additional
dotted the approximation for $M_A \ll 1$ as computed with
(\ref{current2a}).}
\label{f05}
\end{figure}

We use the transformation equation (\ref{alphafromA}) with the Mach number profile
 (\ref{example_MA}) and
$M_{\rm max}=0.5$, $M_{\rm min}=0$, $A_c=2$, and $d=5$ to
derive a stationary equilibrium with plasma flow. The stationary solutions
are shown with dashed lines in figure \ref{f05}. The top panel shows
the flux function $\alpha(y)$, the second panel the magnetic field $B_x$ and the bottom panel
the electric current density $j_z(y)$ as computed with equation (\ref{current1}).
Additionally we present the current approximation for $M_A \ll 1$, as computed with
equation (\ref{current2}) with dotted lines in the bottom panel
of figure \ref{f05}. By comparing the static (solid lines) and
stationary (dashed) quantities, one can see that both quantities only
differ in the region were the flow gradient is high (see third panel).
In these regions, where the static equilibrium is separated from the stationary
flowing plasma in a thin layer, current sheets form. The spatial scale of these
layers is significantly smaller than the typical length scale of the equilibrium
current. The electric current density approximation for small Mach number
(dotted in bottom panel) shows reasonable
agreement with the exact (dashed) solution, even for the not very small
maximum Mach number $M_A=0.5$ used here.
\subsection{Exact 2D magnetospheric equilibrium}
As shown by \citet{liouville1853}, \citet{bandle75}, \citet{birn:etal78}
equation (\ref{lio1}) can be written as:
\begin{equation}
4 \frac{\partial^2 A}{\partial u \,\partial\bar{u}}= \lambda
\exp(-c A) \, , \label{lio1a}
\end{equation}
with $u=x+iy$ and $\bar{u}=x-iy$, and
this equation has the general solution
\begin{equation}
A(u,\bar{u})=\frac{2}{c} \ln \frac{1+\frac{c \lambda }{8}|\Psi(u)^2|}{|\frac{\partial \Psi}
{\partial u}|}\equiv\displaystyle
\hat{A}\ln\displaystyle\frac{1+\displaystyle\frac{1}{4\hat{l}^2}\left|\Psi(u)^2\right|}
{\displaystyle
\left|\frac{\partial \Psi}{\partial u}\right|}
\label{lio2}.
\end{equation}
Every analytic function $\Psi(u)$ generates a solution of (\ref{lio1}). The Liouville equation
has also its applications outside plasma physics and solutions in the form of
(\ref{lio2}) have been used, e.g., by \cite{schmidt-burgk67} to investigate
a self-gravitating gas layer.

\citet{schindler:etal04} found magnetospheric solutions with the Ansatz
\begin{equation}
\Psi(u)=2\hat{l}\,\exp\left( i \left(u/\hat{l}+\sqrt{\frac{u/\hat{l}}{\epsilon}} \right) \right)
\, , \end{equation}
which leads to the solution class of (\ref{lio2}) of
the form \citep[see][for details and discussion of
the static equilibrium]{schindler:etal04}.
\begin{equation}
A(x,y)/\hat{A}=\ln\left(
\frac{\cosh \left(\frac{y}{\sqrt{2 \epsilon} \sqrt{r+x}} +y  \right)}
{\sqrt{\frac{1}{r}\left( \frac{1}{4 \epsilon}+\sqrt{\frac{r+x}{2 \epsilon}}\right)+1}}
\right)
\end{equation}
with $r=\sqrt{x^2+y^2}$, where the coordinates here are normalized on $\hat{l}$.
For $\epsilon\rightarrow \infty$ the Ansatz for $\Psi$ produces the
Harris-sheet solution.
The top panel of figure \ref{f06} shows the corresponding magnetic field lines
as equi-contour plots of $A(x,y)$ for $\epsilon=1$.
For the transformation we used the parameters
$M_1=0.0$, $M_2=0.95$ and $d=2$ for the Mach number profile in Eq.(\ref{example_MA}).
The second panel contains the equilibrium current density $-j_z(x,y)$.
\subsubsection{Transformation to stationary states}
\begin{figure}
\vspace*{2mm}
\begin{center}
\includegraphics[width=5.0cm]{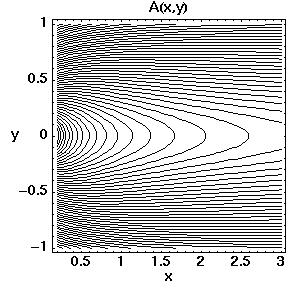} \\ 
\vspace*{3mm}
\includegraphics[width=6.0cm]{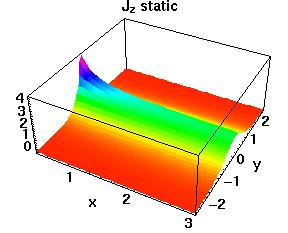} \\ 
\vspace*{3mm}
\includegraphics[width=6.0cm]{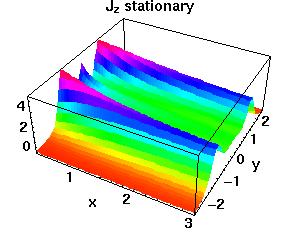} \\ 
\vspace*{3mm}
\includegraphics[width=6.0cm]{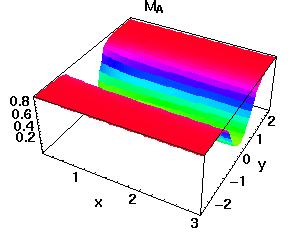} \\ 
\end{center}
\caption{Same transformation formula as figure \ref{f02}, but for a nonlinear
current in the form $J_z(A)\propto \exp(- c A)$.
For the transformation we used $M_1=0.0$, $M_2=0.95$ and $d=2$. For a better visualization we
show the negative current $-j_z(x,y)$ in the second and third panel.
}
\label{f06}
\end{figure}

We use the transformation equation (\ref{alphafromA}) with the Mach number profile
 (\ref{example_MA}) and
$M_{\rm max}=0.95$, $M_{\rm min}=0$, $A_c=0$, and $d=2$ to
derive a stationary equilibrium with plasma flow. The chosen
profile for the flow is smoother, as in previous examples and the
maximum Mach number is higher. As a consequence we observe two
additional current sheets in the stationary current distribution
as computed with (\ref{current2}) and shown in the third panel
of figure \ref{f06}. The thickness of the two current sheets are
located in the region with the plasma flow gradient, shown
in the bottom panel. The smoother profile of $M_A(A)$ results
also in smoother induced current sheets.
\conclusions
\label{conclusions}
Within this work we studied the relation between plasma flow gradients and current sheets
in space plasma. To highlight the influence of stationary
flows on static MHD equilibria, we neglected compressibility effects and
used the assumption of field-aligned, incompressible stationary
flows. These assumptions imply an analogy between magnetic field and
velocity field as well as an analogy between MHS and
incompressible SMHD:
The assumption of incompressibility allows us to transform magnetostatic equilibria
into stationary ones by using a non-canonical transformation.
We find that the occurrence of flow driven current sheets is closely related to
the gradient of the plasma flow or, to be precise, to the gradient of the Alfv\'en Mach number
perpendicular to the magnetic field lines. Along the field lines the Mach number is
always constant for incompressible stationary flows.
As the gradients in the Alfv\'en Mach number can be very large, because the
typical length scale of the flow is smaller than the length scale of the magnetic
field, the occurence of current sheets is correlated with the
appearance of vortex sheets. Such configurations
can be closely connected to local breakdowns of the frozen-in-flux theorem, or shortly
to magnetic reconnection \citep{eyink06}. Magnetic reconnection in turn plays a major
role in eruptive space plasma processes like magnetospheric substorms or solar flares.

In principle it is possible to compare our theoretical investigations on the relation
between plasma flows and current sheets with observations, in particular in the magnetosphere
where in-situ measurements are available. One possibility is using magnetic field and particle
data from the CLUSTER-mission, which are carried out simultaneously with four spacecraft. Such
multi-spacecraft measurements (with distances between the spacecraft in the
range of about 50-10000 km) of the magnetic field allow also the estimation of electric currents.
By taking moments of the particle data it is possible to compute plasma quantities like
density, pressure and the plasma  flow velocity. These combined measurements allow at
least to estimate gradients in the plasma flow and the thickness of
current sheets. A limitation is that structures smaller than the distance of the
Cluster-spacecraft cannot be spatially resolved, which implies that flow gradients could be steeper
and the current sheets thinner as computed from the measurements. A comparison of data with our
model, which relates flows and flow gradients to current sheets, will allow to investigate how
consistent different areas in the magnetosphere can be described under the assumption of
stationary incompressible MHD.

%
%
\begin{acknowledgements}
D.H.N. acknowledges financial support from GAAV \v{C}R under grant number
IAA300030804. D.H.N. is also grateful to the Max-Planck-Institut f\"{u}r Sonnensystemforschung
for the financial support during his visits.
The work of T.W. was supported by DLR-grant 50 OC 0501.
\end{acknowledgements}


\end{document}